\documentclass[twocolumn,showpacs,fleqn,nobibnotes]{revtex4}

\usepackage{amsmath}
\usepackage{graphicx}
\usepackage{float}
\usepackage{subfigure}

\newcommand{\xgrv}{x_{\text{GRV}}}

\begin{document}

\title{Regge residues from DGLAP evolution}
\author{G. Soyez}
\email{g.soyez@ulg.ac.be}
\affiliation{Inst. de Physique, B\^{a}t. B5, Universit\'{e} de Li\`{e}ge, Sart-Tilman, B4000 Li\`{e}ge, Belgium}

\begin{abstract}
We show that combining forward and backward evolution allows to extract the residues of the triple-pole pomeron and of the other singularities for $10$ GeV$^2 \le Q^2 \le 1000$ GeV$^2$. In this approach, the essential singularity generated by the DGLAP evolution is considered as a numerical approximation to a triple-pole pomeron. Using an analytical expression for the form factors, we reproduce the experimental data with a $\chi^2/dof$ of 1.02. This proves the compatibility between Regge theory and DGLAP evolution. The method used here enables us to evaluate the uncertainties on the gluon distribution which prove to be large at small $x$ and small $Q^2$.
\end{abstract}

\maketitle

\section{Introduction}\label{sec:intro}

Deep Inelastic Scattering (DIS) is a very important tool to study the fundamental behaviour of strong interactions. The numerous available data, especially the last ones from HERA, which have very small error bars, allow for a direct test of theoretical models.

The work presented here puts together two major theoretical approaches: perturbative QCD (pQCD), through the Dokshitzer-Gribov-Lipatov-Altarelli-Parisi (DGLAP) evolution and analytical $S$-matrix theory, through Regge theory, on the other. Models using one or the other are numerous. On the one hand, there are a lot of well-known DGLAP fits to experimental data \cite{Martin:2001es, Gluck:1998xa, Pumplin:2002vw, Cooper-Sarkar:2002yx, Alekhin:2000ch} which reproduce very well the experimental data. However, in the usual parton distribution sets, each initial parton distribution presents its own singularities. As an example, in the MRST2001 parametrization \cite{Martin:2001es}, we have
\[
xq(x, Q_0^2) = A (1+B\sqrt{x}+Cx)(1-x)^{\eta_q}x^{\varepsilon_q},
\]
with $\varepsilon_\text{sea}=-0.26$, $\varepsilon_g=-0.33$. Such a behaviour is not consistent with Regge theory.
On the other hand, Regge theory establishes strong constraints on the large $s$ (small $x$) behaviour of total cross sections, and Regge-compatible fits to experimental data are also numerous \cite{Cudell:2002ej, Cudell:2001ii, Desgrolard:2001bu, Donnachie:2001xx, compete}. In those fits, the residues of the Regge poles are unknown functions of $Q^2$, and one can only predict $F_2$ while quarks and gluon distribution functions remain out of reach.

In a previous paper \cite{Soyez:2002nm}, we have shown that one can use a triple-pole pomeron to parametrize the initial parton distribution functions and let them evolve according to the DGLAP equation. In this approach, both quarks and gluons have the same singularity structure. Our model is therefore compatible both with Regge theory and with pQCD. In addition, we predict a gluon distribution of the same order of magnitude as the one obtained in the usual DGLAP fits.

We now want to explore whether it is possible to avoid the essential singularity coming from DGLAP evolution and extend the Regge fit to large $Q^2$ values. In other words, can we use the DGLAP evolution equations to find the $Q^2$ dependence of the Regge form factors ? To answer that question, the main obstacle lies in the fact that the DGLAP evolution generates an essential singularity which seems to be incompatible with the Regge ones. To solve that problem, we need to remember that the DGLAP evolution is only an approximation and that only a full resummation of the perturbative series - including all order corrections to DGLAP, small-$x$ and higher-twist effects - should give the correct analytical behaviour. From the large number of DGLAP fits, one expects corrections to be small and DGLAP to be a good approximation. We will therefore choose a Regge compatible singularity structure and an initial scale $Q_0^ 2$ for DGLAP evolution, and then adjust the initial distribution in order to reproduce the experimental data as well as possible. In this way, we obtain the residues at scale $Q_0^2$ and, repeating this procedure for different values of $Q_0^2$ we get the Regge form factors for the full range of the initial scale. In this method, it is important to note that we keep a common singularity structure, namely a triple-pole pomeron and an $f,a_2$-reggeon, in all parton distributions. 

The structure of the paper is as follow: in section 2, we introduce the basics of DGLAP evolution and Regge theory, we emphasise the tools needed in the rest of the paper and we summarise previous work. In section 3, we give the form of the initial distribution that we will use as initial condition for DGLAP evolution and that is obtained from Regge theory. Section 4 tells how we use DGLAP evolution to obtain the form factors and section 5 gives the results of the fit. We conclude in section 6.

\section{Perturbative QCD and Regge theory}\label{sec:theo}

\subsection{Perturbative QCD}\label{sec:pqcd}

In pQCD, the high-$Q^2$ behaviour of DIS is given by the DGLAP evolution equations \cite{DGLAP}. These equations introduce the {\em parton distribution functions} $q_i(x, Q^2)$, $\bar q_i(x, Q^2)$ and $g(x, Q^2)$, which can be interpreted as the probability of finding, in the proton, respectively a quark, an anti-quark or a gluon with virtuality less than $Q^2$ and with  longitudinal momentum fraction $x$. When $Q^2 \to \infty$, the $Q^2$ evolution of these densities (at fixed $x$) are given by the DGLAP equations
\begin{eqnarray}\label{eq:DGLAP}
\lefteqn{ Q^2\partial_{Q^2}
\begin{pmatrix}q_i(x,Q^2)\\\bar q_i(x,Q^2)\\g(x,Q^2)\end{pmatrix}}\\
& = &\frac{\alpha_s}{2\pi} \int_x^1 \frac{d\xi}{\xi}
\left.\begin{pmatrix}
 P_{q_iq_j} & . & P_{q_ig}\\
 . & P_{q_iq_j} & P_{q_ig}\\
 P_{gq} & P_{gq} & P_{gg}
\end{pmatrix}\right|_{\frac{x}{\xi}}
\begin{pmatrix} q_j(\xi,Q^2)\\\bar q_j(\xi,Q^2)\\ g(\xi, Q^2) \end{pmatrix}.\nonumber
\end{eqnarray}
At leading order, $P_{xy}$ does not depend on $Q^2$. The connection to $F_2$ is given by
\begin{equation}\label{eq:F2}
F_2(x,Q^2) = x \sum_i e_{q_i}^2 \left[q_i(x,Q^2) + \bar q_i(x,Q^2) \right],
\end{equation}
where the sum runs over all quark flavours.

The usual way to use this equation is to choose a set of initial distributions $q_i(x, Q_0^2, \vec{a})$ to compute $q_i(x, Q^2, \vec{a})$ using \eqref{eq:DGLAP} and to adjust the parameters $\vec{a}$ in order to reproduce the experimental data.
This approach have already been successfully applied many times \cite{Martin:2001es, Gluck:1998xa, Pumplin:2002vw, Cooper-Sarkar:2002yx, Alekhin:2000ch} and is often considered a very good test of pQCD.
Nevertheless, these studies do not care about the singularity structure of the initial distributions, ending up with results that disagree with Regge theory, and presumably with QCD.

\subsection{Regge theory}\label{sec:regge}

Beside the predictions of pQCD, we can study DIS through its analytical properties. In Regge theory \cite{Regge:mz,Regge:1960zc}, we consider amplitudes ${\cal A}(j,t)$ in the complex angular momentum space by performing a Sommerfeld-Watson transform. In that formalism, the large-$s$ dependence of amplitudes is determined by their singularity structure in $j$-plane. The singularities are functions of $t$ only and this technique can be applied to the domain $\cos(\theta_t) \gg 1$, where $\theta_t$ is the deviation angle in the center-of-mass frame for the $t$-channel amplitudes and where amplitudes are analytically continued to any $\cos(\theta_t)$ value. For example, we can fit the DIS data or the photon structure function at large $p.q\equiv \nu$ (small $x$), and the total cross sections at large $s$. 

The models based on Regge theory \cite{Cudell:2001ii,Desgrolard:2001bu,Donnachie:2001xx,books} use a pomeron term, reproducing the rise of the structure function (resp. cross sections) at small $x$ (resp. large $s$), and reggeon contributions coming from the exchange of meson trajectories ($a$ and $f$). 

We shall consider here the following parametrization for the pomeron contribution to $F_2$ or to $\sigma_{\text{tot}}$:
\begin{equation}\label{eq:triple}
a(Q^2)\ln^2\left[\nu/\nu_0(Q^2)\right] + c(Q^2),
\end{equation}
corresponding to a triple pole in the complex $j$-plane \cite{Cudell:2002ej}
\[
\frac{a}{(j-1)^3}-\frac{2a\ln(\nu_0)}{(j-1)^2}+\frac{a\ln^2(\nu_0)+c}{j-1}.
\]
%\end{itemize}
This seems to be the preferred phenomenological choice at $Q^2=0$ \cite{compete}.
Note that the upper expression, given in terms of $\nu$ and $Q^2$, can be rewritten in terms of $Q^2$ and $x=Q^2/(2\nu)$. So, the $x$ behaviour of the structure functions is constrained by Regge theory.

However, this approach has two problems:
\begin{enumerate}
\item it only predicts $F_2$, while a full set of parton distribution functions is needed,
\item it does not give any information about the $Q^2$ dependence of the residues.
\end{enumerate}

\section{Mixing DGLAP evolution with Regge theory}

From a previous study \cite{Soyez:2002nm}, we have learned that the small-$x$ behaviour of the parton distributions can be described by a triple-pole pomeron model at the initial scale $Q_0^2$ and then evolved using the DGLAP equation. We can summarise these results as follows:
\begin{itemize}
\item \textit{It is possible to have the same singularity structure for quarks and gluons, in agreement with Regge theory}. 
\item One can use a triple-pole pomeron to parametrize the initial parton distributions. In particular, we predict a gluon distribution compatible with usual ones.
\item The DGLAP evolution generates an essential singularity at $j=1$. We expect that this singularity is purely of perturbative origin and that a full resummation of QCD should lead to a Regge-compatible singularity structure.
\item The $F_2$ residues were fixed to reproduce results from the soft QCD fit \cite{Cudell:2002ej} which goes down to $Q^2=0$.
\end{itemize}

We show here that we can use this trick to obtain form factors which are compatible with DGLAP evolution. We present in this section the model used to extract these residues as a function of $Q^2$. In summary, always considering that DGLAP evolution does not lead to a relevant singularity structure but only to an approximation of the parton distributions, we consider that the initial distribution fitted by starting the evolution at the scale $Q_0^2$ gives the residues at $Q_0^2$. We just need to repeat this by varying $Q_0^2$ to get the residues over the whole $Q^2$ range. Note that since we want to {\em predict} the $F_2$ residues, we do not constrain $F_2$ with a soft QCD fit as it was done in \cite{Soyez:2002nm}. However, it is known that the triple-pole pomeron model can be used to reproduce soft processes. Therefore, one knows that the residues found here can be extended down to $Q^2=0$. Since we only want to study the domain where both DGLAP and Regge theory are expected to apply, we won't go into such an extension.

%\subsection{The large-$x$ problem}

Regge theory is expected to be valid for $\cos(\theta_t) \gg 1$. In the case of DIS, $\cos(\theta_t) = \frac{\sqrt{Q^2}}{x\,m_p}$, where $m_p$ is the mass of the proton. The Regge region is thus
\begin{equation}\label{eq:ReggeLim}
\frac{\sqrt{Q^2}}{x\,m_p} \ge K,
\end{equation}
with $K$ a fixed number.
We clearly see that the domain does not extends up to $x=1$. However, in the DGLAP evolution equations \eqref{eq:DGLAP} the small-$x$ domain is coupled to large-$x$ values and therefore we need a parametrization for the parton distribution functions at large $x$. The ideal case is certainly the introduction of powers of $(1-x)$, or of any polynomial vanishing at $x=1$, in our initial distributions. This should take into account daughter trajectories in Regge theory and ensure that parton distributions go to 0 as $x \to 1$. Unfortunately, this has two drawbacks:
\begin{enumerate}
\item it introduces a lot of additional parameters which are not directly related to the small-$x$ Regge behaviour,
\item a precise description of the parton distributions at large $x$ requires more than simply fitting $F_2^p$ and splitting it into flavor singlet component, coupled to the gluons, and one additional non-singlet distribution. We need to introduce valence quark distributions and sea quark distributions. Fitting many kinds of experiments like $F_2^p$, $F_2^n$, $F_2^d$, $F_2^{\nu N}$ and $xF_3^{\nu N}$ allows to constraint all these distributions separately. Again, introducing more parton distributions requires more parameters. Moreover, most of the experimental points except those from $F_2^p$, lies outside the region of interest \eqref{eq:ReggeLim}.
\end{enumerate}
In order to keep the parameters of our model as closely linked as possible to the Regge domain of the evolution, we will use an external distribution for the large-$x$ parton distributions. As done in \cite{Soyez:2002nm}, we will use GRV98 \cite{Gluck:1998xa} at large $x$. We have already argued that the results do not depend much on that choice and that their leading singularity structure is not affected by the parametrization at large $x$. Practically, we shall use the GRV parametrization for $x \ge \xgrv$ and our analytical form at small $x$.

%Removing this dependence on an external model and extending our initial condition up to $x=1$ is left for future work.

\begin{figure}[t]
\includegraphics{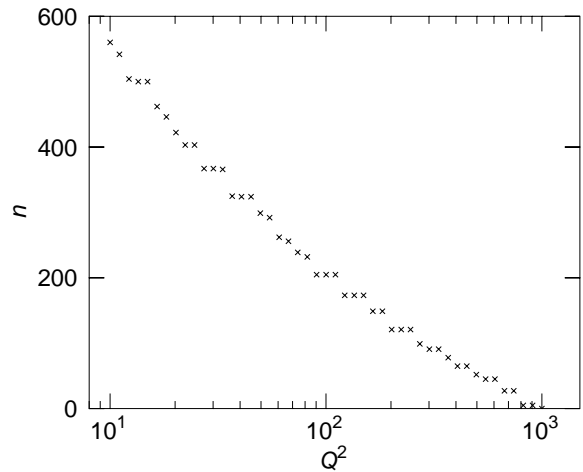}
\caption{Number of points involved in the forward evolution}\label{fig:pts}
\end{figure}

\section{Initial distributions}\label{sec:distrib}

In order to reproduce $F_2^p$, we do not need the full set $\{q_i, \bar q_i, g\}$ of parton distributions. One can quickly check that the minimal set of distributions required is
\begin{eqnarray}
T      & = & x\left\lbrack (u^+ +c^++t^+)-(d^++s^++b^+)\right\rbrack,\\
\Sigma & = & x\left\lbrack (u^+ +c^++t^+)+(d^++s^++b^+)\right\rbrack,\\
G      & = & x g,
\end{eqnarray}
where $q^+ = q+\bar q$ for $q=u,d,s,c,t,b$. The $T$ distribution is a flavor non-singlet distribution, its $Q^2$ evolution is decoupled from any other distribution and is given by a DGLAP equation with $xP_{qq}(x)$ as splitting function. The sea quark distribution $\Sigma$ is flavor singlet and evolves coupled to the gluon distribution using the full splitting matrix. 

Using these distributions, the proton structure function becomes
\[
F_2=\frac{5\Sigma+3T}{18}.
\]

In order to perform a DGLAP evolution, we must find an analytical expression for the $x$ dependence of these distributions at a scale $Q_0^2$, where evolution will start. Since we want to describe $F_2$ as a triple-pole pomeron and a reggeon ($f,a_2$-trajectories), and since adding new singularities in the parton distributions disagrees with Regge theory, the $T$, $\Sigma$ and $G$ distributions will be described by:
\begin{eqnarray*}
T(x,Q_0^2) & = & a_T \log^2(1/x) + b_T \log(1/x) + c_T + d_T x^\eta,\\
\Sigma(x,Q_0^2) & = & a_\Sigma \log^2(1/x) + b_\Sigma \log(1/x) + c_\Sigma + d_\Sigma x^\eta,\\
G(x,Q_0^2) & = & a_G \log^2(1/x) + b_G \log(1/x) + c_G + d_G x^\eta.
\end{eqnarray*}

Out of the 12 parameters in the above expressions, some may be removed. First of all, we expect that the pomeron, having vacuum quantum numbers, does not distinguish between quark flavours. This means that it is decoupled from the $T$ distribution and that we may set $a_T = b_T = c_T = 0$. Then, since the reggeon is mainly constituted of quarks, we shall neglect its contribution to the gluon distribution and set $d_G=0$.

In addition, we ask our distributions to be continuous with GRV's at $x=\xgrv$. This is used to fix $d_T$, $c_\Sigma$ and $c_G$.

Finally, we have seen that we obtain much better results if we multiply each distribution by some power of $(1-x)$. This new factor, effectively including daughter trajectories, is expected from Regge theory. As already pointed out, this factor is not directly connected to the small-$x$ domain. One can see that its main effect is to reproduce an inflexion point present around $x\approx 0.01$ in the parton distributions. In order to minimise the number of parameters, we used only two different powers in all distributions.
We thus end with the following functions
\begin{eqnarray*}
T(x,Q_0^2) & = & d_T^* x^\eta (1-x)^{b_2},\\
\Sigma(x,Q_0^2) & = & a_\Sigma \log^2(1/x) + b_\Sigma \log(1/x) + c_\Sigma^* (1-x)^{b_1}\\
                & + & d_\Sigma x^\eta(1-x)^{b_2}\\
G(x,Q_0^2) & = & a_G \log^2(1/x) + b_G \log(1/x) + c_G^* (1-x)^{b_1}.
\end{eqnarray*}
The parameters with a superscript $^*$ are constrained by continuity with GRV parametrisation and 7 free parameters remain: $a_\Sigma$, $b_\Sigma$, $d_\Sigma$, $a_G$, $b_G$, $b_1$ and $b_2$.

These parameters must depend on $Q_0^2$ and we will see in the next section how we can find their $Q^2$ dependence.

\begin{figure}[ht]
\includegraphics{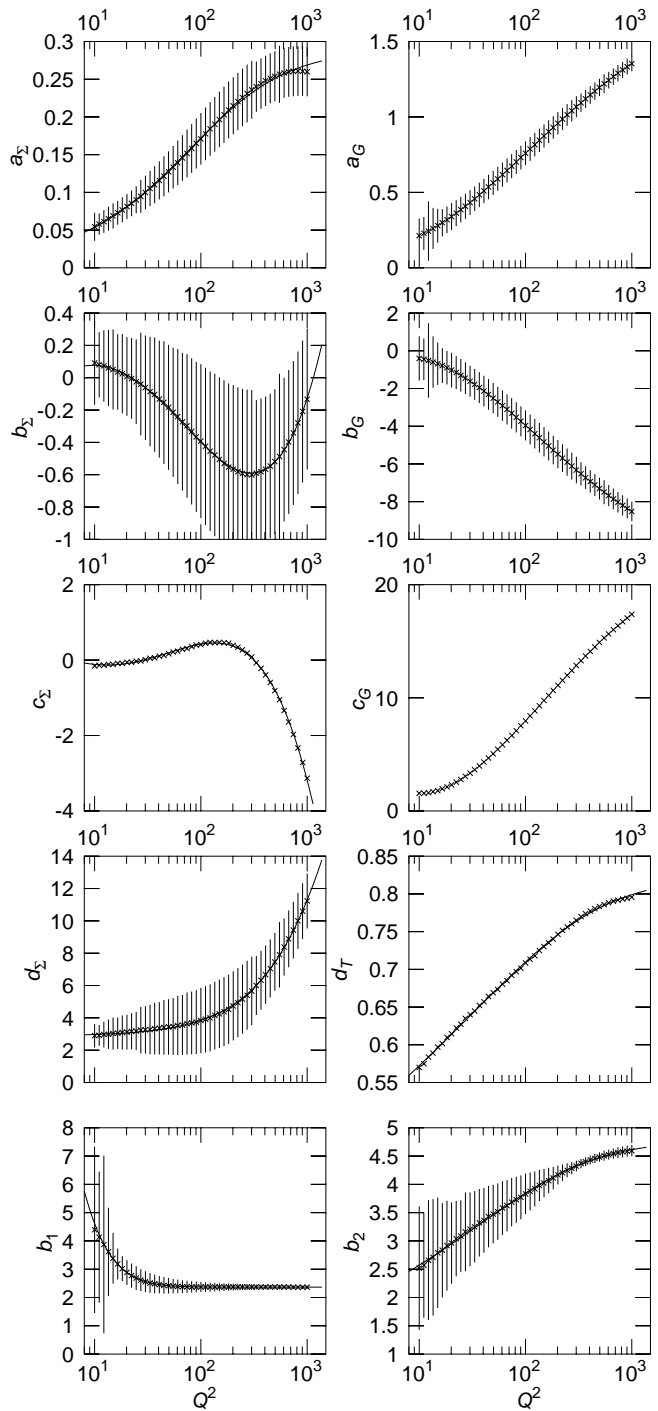}
\caption{Form factors as functions of $Q_0^2$}\label{fig:forms}
\end{figure}

\section{Fitting the form factors}\label{sec:form}

It is well-known that DGLAP evolution generates an essential singularity at $j=1$. From that point of view, finding the $Q^2$ dependence of the residues of a triple-pole pomeron may seem impossible. However, DGLAP evolution always involves a finite order calculation of the splitting functions and, moreover, it resums large $Q^2$ corrections while resumming large $s$ (small $x$) contributions should be preferable for a comparison with Regge theory. Consequently, we argue that the singularity structure generated by DGLAP evolution is physically irrelevant and that only a full resummation of QCD should end up on a correct Regge behaviour.

One must note that if we consider a resumation {\em \`a la} BFKL, we have a stable square-root branchpoint instead of an essential singularity. Furthermore, even in this case we still need to unitarize the pomeron, which may lead to another singularity structure.

So, which singularity do we need to take ? How can we use the DGLAP evolution to calculate residues in Regge theory ? Our point of view is that, although it doesn't provide a correct Regge behaviour, DGLAP evolution can still be considered as a good numerical approximation. In such a case, the only place where Regge theory can be used is for the initial distribution. We end up with a Regge-compatible distribution at the initial scale $Q_0^2$ and a DGLAP approximation everywhere else.

As we shall see, obtaining the Regge residues from this technique is a quite straightforward.

\subsection{Scheme 1}\label{sec:form1}

As initial condition for DGLAP evolution, we shall use the distributions obtained in section \ref{sec:distrib} for $x\le \xgrv$ and the GRV98 distribution for larger $x$ values. The parameters are considered as functions of $Q^2$ and they can be found as follows:
\begin{enumerate}
\item\label{s11} choose an initial scale $Q_0^2$,
\item\label{s12} choose a value for the parameters in the initial distribution,
\item\label{s13} compute the parton distributions at larger $Q^2$ using DGLAP evolution equations,
\item\label{s14} repeat \ref{s12} and \ref{s13} until you find the best values of the parameters reproducing the experimental $F_2$ data for $Q^2>Q_0^2$ and $x\le \xgrv$.
\item This gives the residues at the scale $Q_0^2$ and we repeat steps \ref{s11} to \ref{s14} in order to obtain the residues at all $Q^2$ values.
\end{enumerate}

This technique works quite well but has one drawback: calculation of the residues at the scale $Q_0^2$ relies on the experimental points satisfying
\[
Q_0^2 \le Q^2 \le Q_{\text{max}}^2.
\]
Unfortunately, this domain depends on the initial scale and as we can see on Fig. \ref{fig:pts} 
the number of points used for the fit decrease quite fast when $Q_0^2$ rises.

\subsection{Scheme 2}\label{sec:form2}

There is a way to keep the same set of points in each fits: adding backward evolution. The previous scheme thus becomes 
\begin{enumerate}
\item\label{s21} choose an initial scale $Q_0^2$,
\item\label{s22} choose a value for the parameters in the initial distribution,
\item\label{s23} compute the parton distributions for $Q_0^2 \le Q^2  \le Q_{\text{max}}^2$ using forward DGLAP evolution and for $Q_{\text{min}}^2 \le Q^2  \le Q_0^2$ using backward DGLAP evolution,
\item\label{s24} repeat \ref{s22} and \ref{s23} until you find the value of the parameters reproducing the $F_2$ data for $Q^2>Q_{\text{min}}^2$ and $x\le \xgrv$.
\item This gives the residues at the scale $Q_0^2$ and we repeat steps \ref{s21} to \ref{s24} in order to obtain the residues at all $Q^2$ values.
\end{enumerate}
This way, we take into account the experimental points with $Q_{\text{min}}^2 \le Q^2  \le Q_{\text{max}}^2$ whatever the value of $Q_0^2$ is.

\begin{figure}
\includegraphics[scale=0.67]{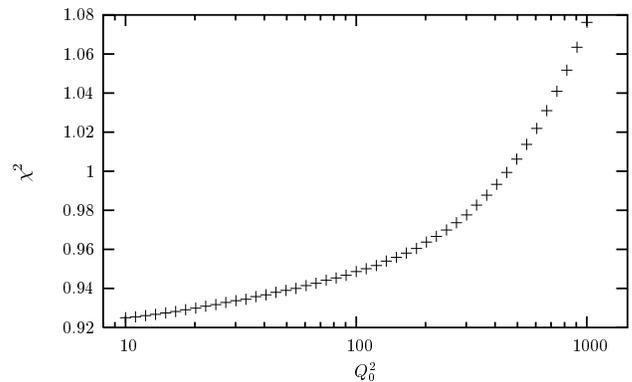}
\caption{$\chi^2/\text{n}$ of the fit as a function of the initial scale for DGLAP evolution}\label{fig:chi2-ud}
\end{figure}

\section{Results}\label{sec:results}

\begin{table}
\begin{tabular}{lccc}
\hline
\hline
Experiment & nb. pts. & $\chi^2$ & $\chi^2/n$ \\
\hline
BCDMS\cite{Benvenuti:1989rh} &   5 &  20.9 & 4.185\\
E665\cite{Adams:1996gu}      &  10 &   8.0 & 0.797\\
H1\cite{Adloff:1999ah,Adloff:2000qj,Adloff:2000qk}    & 240 & 219.2 & 0.913\\
NMC\cite{NMC}   &  11 &  17.1 & 1.553\\
ZEUS\cite{Breitweg:1999ad,Chekanov:2001qu}  & 294 & 306.3 & 1.042\\
\hline
Total & 560 & 571.5 & 1.020\\
\hline
\hline
\end{tabular}
\caption{Experimental points in the domain \eqref{eq:domain} and results of the analytical triple-pole fit}\label{tab:chi2}
\end{table}

We have chosen to use the second scheme with both forward and backward evolution. We have fitted the data in the region 
\begin{equation}\label{eq:domain}
\begin{cases}
10 \le Q^2 \le 1000\:\text{GeV}^2, \\
x\le \xgrv = 0.15,\\
\cos(\theta_t) = \frac{\sqrt{Q^2}}{xm_p} \ge \frac{49\:\text{GeV}^2}{2m_p^2}.
\end{cases}
\end{equation}
This method ensures that $Q^2$ is large enough to apply DGLAP evolution, that $\cos(\theta_t)$ is large enough to use Regge theory and that the large-$x$ domain is excluded from the fit. The limit on $\cos(\theta_t)$ has been taken from the pure Regge fit of \cite{Cudell:2002ej}. In this region, the experimental points are as shown in Table \ref{tab:chi2}.

\subsection{Form factors}

When we apply the method explained in section \ref{sec:form2} using the parametric distributions obtained in section \ref{sec:distrib}, we obtain the form factors presented in Fig. \ref{fig:forms} and calculated from 10 to 1000 GeV$^2$. The $\chi^2$ per data point obtained for each fit, shown in Fig. \ref{fig:chi2-ud}, is of order 1 and the form factors appear to be smooth functions of $Q_0^2$.

%One can see that the $\chi^2$ of the fit grows slightly when $Q_0^2$ is very large. This may be understood as follows: we used DGLAP equations at leading order (LO), thus the corrections introduced by moving to NLO evolution should behave like 
%\[
%\frac{\alpha_s(Q^2)}{2\pi} \sim \frac{1}{\log(Q^2/\Lambda_{QCD}^2)}.
%\]
%Since these corrections become more and more important at low $Q^2$, we expect backward evolution to be less precise than forward evolution. In other words, backward evolution is less stable under higher order corrections. When $Q_0^2$ is large, we evolve mostly backward and therefore the $\chi^2$ grows.
%Due to this growth, it is difficult to go to smaller values of $Q^2$ and keep an acceptable $\chi^2$ for large $Q_0^2$ values. One must also point out that DGLAP evolution couples small-$x$ values to larger ones. Therefore, even when $Q_0^2$ is large, our initial distribution must go down to $x\approx 10^{-4}$, while there is no experimental data below $x\approx 0.01$ at large $Q^2$. These distributions extend to values of $\nu$ much higher than the experimental ones and in addition to DGLAP corrections, one might also expect BFKL corrections. Nevertheless, this seems to prove that the triple-pole pomeron model can be extended to higher values of $\nu$.

Finally, when choosing $\xgrv$, we do not want to take the high-$x$ experimental points for reasons explained previously, but we want to set $\xgrv$ as high as possible to maximize the number of experimental point as possible, particularly in the large $Q^2$ domain where DGLAP evolution is expected to work better. We found that $\xgrv=0.15$ was a good compromise between those constraints.

\begin{figure*}
\includegraphics[scale=0.9]{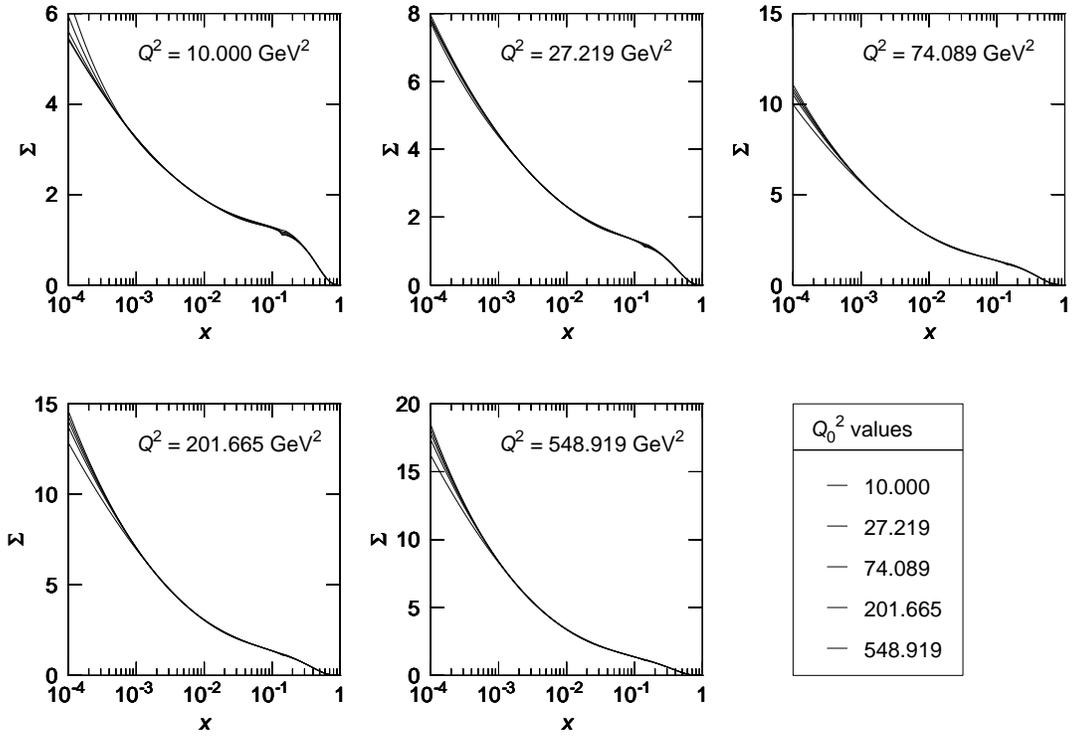}
\caption{$\Sigma$ distribution at various $Q^2$. Different curves correspond to different values of $Q_0^ 2$.}\label{fig:pdfS}
\end{figure*}

\begin{figure*}
\includegraphics[scale=0.9]{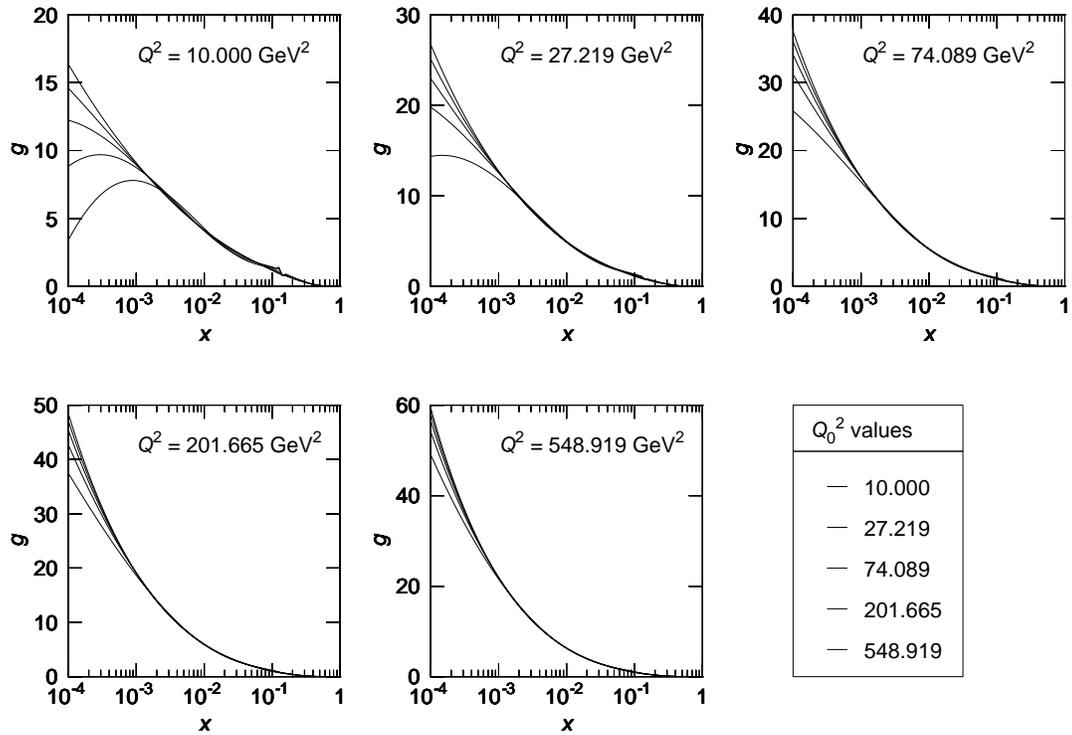}
\caption{gluon distribution at various $Q^2$. Different curves correspond to different values of $Q_0^ 2$.}\label{fig:pdfG}
\end{figure*}

\subsection{Parton Distributions}

Once we have fitted the form factors, it is interesting to see if the distributions we obtain at a scale $Q_0^2$ are the same if we start evolution at $Q_0^2$ or if we start at $Q_1^2$ and evolve until $Q_0^2$. As we see in Fig. \ref{fig:pdfS}, for the sea-quark distribution, the difference is very small. This shows clearly that the DGLAP essential singularity can be considered as a numerical approximation to a triple-pole singularity, at least for quarks.

For the gluon case, presented in Fig. \ref{fig:pdfG}, things are different. One can see that backward evolution tends to produce lower gluon distribution at small $x$. These can even become negative if we evolve to lower $Q^2$ values. However, the structure function only depends on the quark distributions which is stable as shown in Fig. \ref{fig:pdfS}. Thus, from the phenomenological point of view, all these distributions provide a correct description of the data and are perfectly acceptable. The differences in the gluon distribution obtained for different values of $Q_0^2$ must therefore be considered as uncertainties on the gluon distribution. One clearly sees from Fig. \ref{fig:pdfG} that even at $Q^2=10$ GeV$^2$, the uncertainties are very large. These large errors on the gluon distribution at small $x$ and small $Q^2$ can be of prime importance in the study of LHC physics.

\subsection{Analytical form factors}

\begin{figure*}
\includegraphics{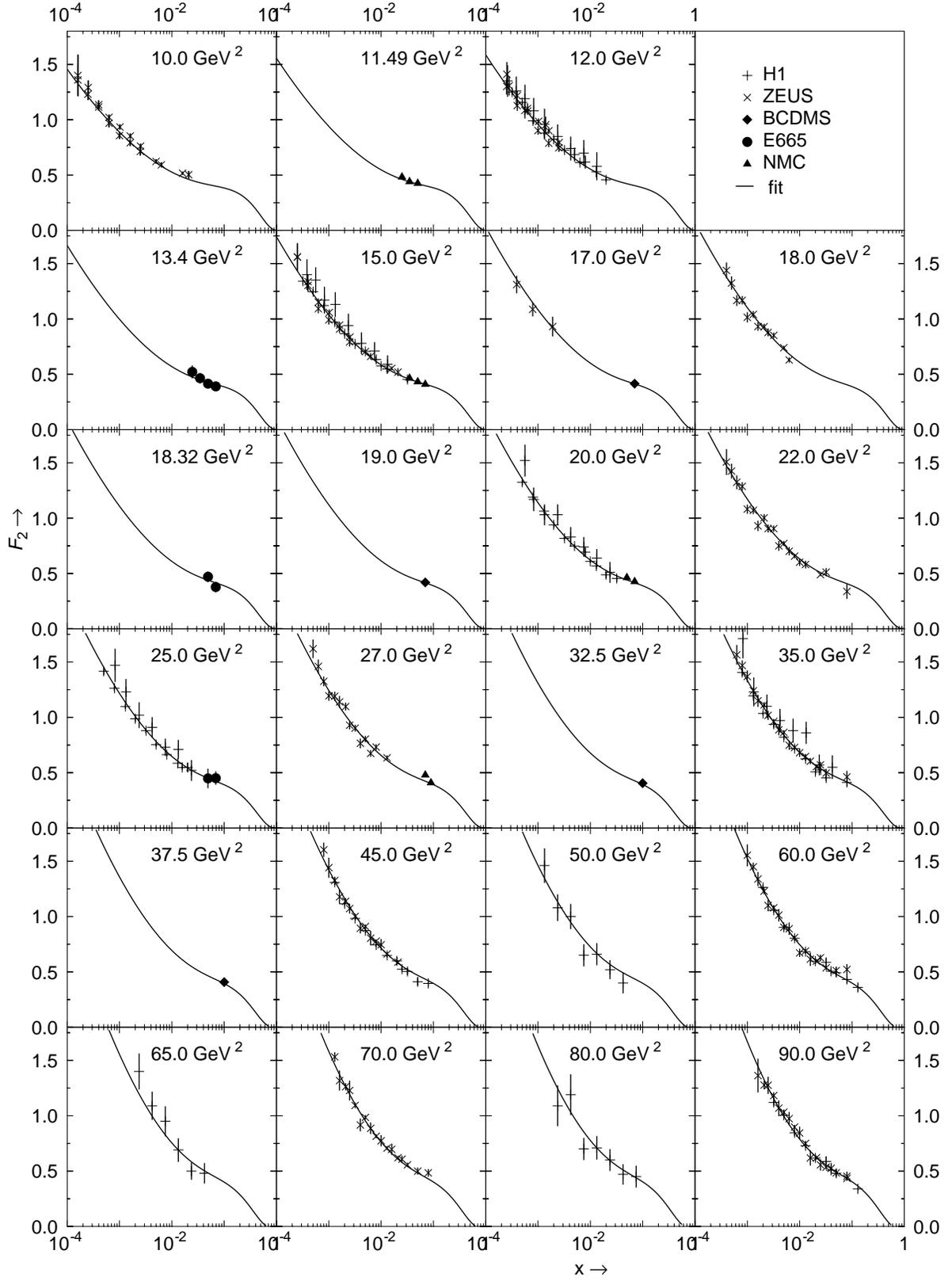}
\caption{$F_2$ fit using analytical approximation of the residues (low $Q^ 2$ values).}\label{fig:lowq2}
\end{figure*}

\begin{figure*}
\includegraphics{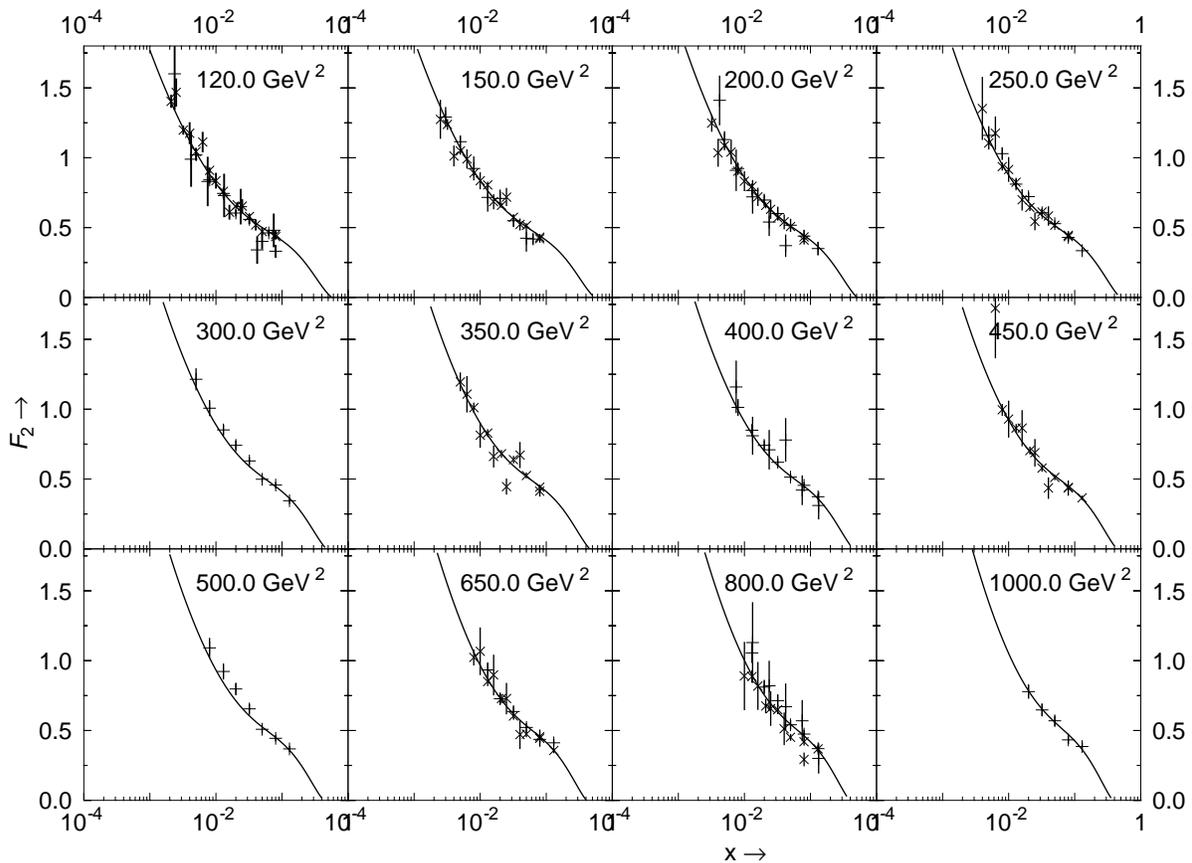}
\caption{$F_2$ fit using analytical approximation of the residues (high $Q^ 2$ values).}\label{fig:highq2}
\end{figure*}

As a final step, we have tried to find analytical expressions for the form factors produced by the fit. Actually, it is much easier to work with analytical expressions reproducing results of Fig. \ref{fig:forms} than to refit everything for each $Q_0^2$ values. In the $Q^2$-range studied here, we have found that the following expressions, plotted in Fig. \ref{fig:forms}, reproduce very well the form factors extracted from the DGLAP fit.

\begin{eqnarray*}
a_\Sigma & = & 0.29 \left(\frac{Q^2}{Q^2+123.3}\right)^{0.65}\\
b_\Sigma & = & \left[0.0048\ln^2\left(\frac{Q^2}{9.666}\right) -0.114\right]\ln^2\left(\frac{Q^2}{9.666}\right) - 0.077\\
c_\Sigma & = & \left[0.218-0.02\ln^2\left(\frac{Q^2}{13.1}\right) \right]\ln^2\left(\frac{Q^2}{13.1}\right) - 0.13\\
d_\Sigma & = & 2.66 \ln^2\left(2.83+\frac{Q^2}{199.13}\right)\\
d_T      & = & 0.82 \left(\frac{Q^2}{Q^2+301.8}\right)^{0.105}\\
b_1      & = & 2.367+\ln^{3.38}\left(1.0+\frac{25.63}{Q^2}\right)\\
b_2      & = & 4.78 \left(\frac{Q^2}{Q^2+181.8}\right)^{0.21}
\end{eqnarray*}

Using these analytical expressions, one can compare their predictions with the experimental data. We obtain a $\chi^2$ of 1.02 per data point, in a very good agreement with experimental measurements. The results are presented in Table \ref{tab:chi2} and in Figs. \ref{fig:lowq2} and  \ref{fig:highq2}.

In the above equations, we only wanted to find some analytical expression reproducing the form factors extracted numerically from the DGLAP fits. We have not taken care of their analytical properties in $Q^2$. To obtain better expressions, one should find an analytical equation for the form-factors $Q^2$-evolution. Since we have seen that the evolved distributions, containing an essential singularity, are numerically close to the triple-pole distributions, one may hope that subleading corrections to the DGLAP equations will stabilize a triple-pole distribution.

\subsection{$F_2^c$ and $F_L$}

\begin{figure*}
\includegraphics[scale=1.0]{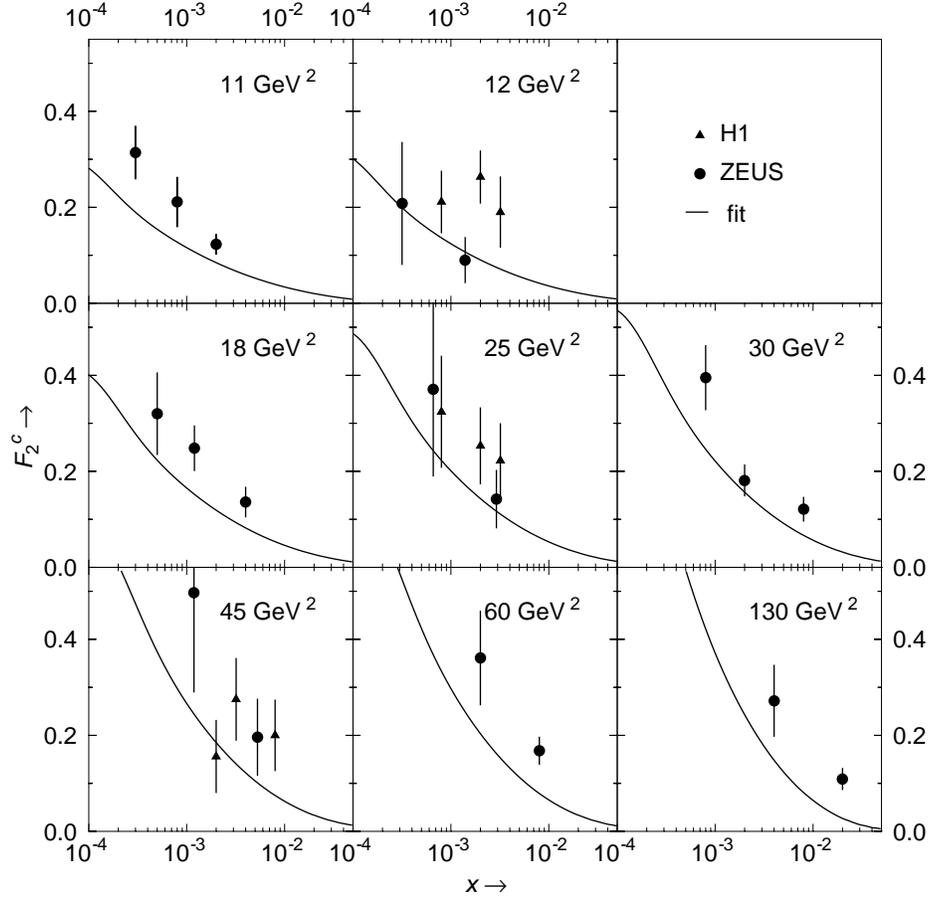}
\caption{Predictions for the charm structure function using our gluon distributions}\label{fig:fc}
\end{figure*}

\begin{figure*}
\includegraphics[scale=1.0]{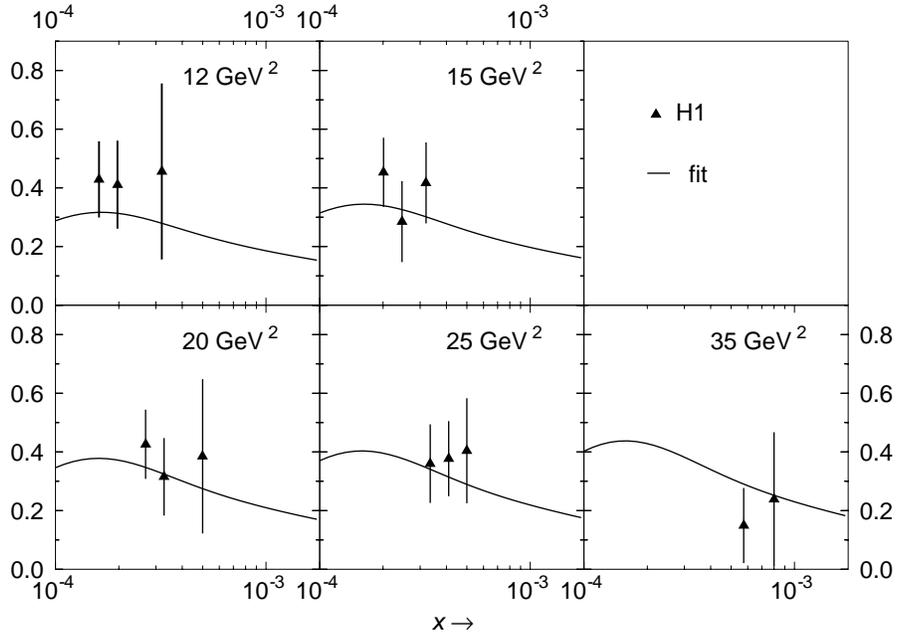}
\caption{Predictions for the longitudinal structure function using our parton distributions}\label{fig:fl}
\end{figure*}

Although only $F_2^p$ is included in the fit, it is interesting to see if our predictions agree with the measurements of $F_2^c$ \cite{Adloff:1996xq,Breitweg:1997mj,Breitweg:1999ad} and $F_L$ \cite{Adloff:2000qk}. Actually, these quantities are directly sensitive to the gluon distribution and are therefore a good test of our predictions. The charm structure function is given by \cite{book:roberts}
\begin{eqnarray*}
F_2^c(x,Q^2) & = & 2e_c^2\frac{\alpha_s(Q^2+4m_c^2)}{2\pi}\\
             &   & \int_{ax}^1 d\xi\,g(\xi, Q^2+4m_c^2)f(x/\xi, Q^2),
\end{eqnarray*}
with
\begin{eqnarray*}
f(x,Q^2) & = & v\left\lbrack(4-\mu)x^2(1-x)-\frac{x}{2}\right\rbrack \\
         & + & L \left\lbrack \frac{x}{2}- x^2(1-x)+\mu x^2(1-3x)-\mu^2z^3\right\rbrack,\\
\mu & = & \frac{2m_c^2}{Q^2}, \\
v & = & \sqrt{1-\frac{2x\mu}{1-x}},\\
L & = & \log\left(\frac{1+v}{1-v}\right)\\
a & = & 1+2\mu.
\end{eqnarray*}
We have adopted a value of 1.3 GeV for the charm quark mass and the predictions for $F_2^c$ obtained from our model are presented in Fig. \ref{fig:fc}. We see that the data are quite well reproduced.

For the case of the longitudinal structure function, we are sensitive both to quarks and gluons:
\[
F_L = \sum_{q=u,d,s,c} G_q + \frac{4\alpha_s(Q^2)}{3\pi}\int_x^1\frac{d\xi}{\xi}\,\left(\frac{x}{\xi}\right)^2F_2(\xi,Q^2),
\]
where
\begin{eqnarray*}
G_q & = & 2e_q^2\frac{\alpha_s(Q^2+4m_q^2)}{\pi}\int_{ax}^1 d\xi\,\left(\frac{x}{\xi}\right)^2 g(\xi, Q^2+4m_c^2)\\
    &   & \phantom{2e_q^2\frac{\alpha_s(Q^2+4m_c^2)}{\pi}\int_{ax}^1 d\xi}\left\lbrack \left(1-\frac{x}{\xi}\right)v-\frac{\mu x}{\xi}L\right\rbrack
\end{eqnarray*}
with $m_u=m_d=m_s=0$ and $m_c=1.3$ GeV. The results obtained from our parton distributions set are shown in Fig. \ref{fig:fl} and show good agreement with the data.

\section{Conclusion}\label{sec:ccl}

In summary, we have seen in this paper that DGLAP evolution can be used to extract the residues of a triple-pole pomeron model. We chose triple-pole distributions for both quarks and gluons at an initial scale $Q_0^2$ and we evolved them using DGLAP evolution. A comparison with the experimental data gives the residues of the triple-pole pomeron at the scale $Q_0^2$. In order to keep the same data points when $Q_0^2$ varies, we evolved the initial distribution both forward and backward between $Q_{\text{min}}^2$ and $Q_{\text{max}}^2$. We have repeated that fit for various values of $Q_0^2$ resulting in smooth form factors for the triple-pole residues. 

In this model, quarks and gluons have the same singularity structure, as suggested by Regge theory. The essential singularity generated by DGLAP evolution appears to be a good numerical approximation to a triple pole for the $F_2$ structure function. For the gluons, we found that there were quite large uncertainties in the gluon distribution at small $x$ and small $Q^2$. These uncertainties should certainly be taken into account in the LHC phenomenology.

Finally, we were able to find analytical expressions for our form factors. A comparison of these expressions with experimental data gives a $\chi^2$ per point of 1.02, which proves the very good agreement between pQCD and Regge theory. We have also checked that our gluon distribution was acceptable by comparing our predictions for $F_2^c$ and $F_L$ with the H1 and ZEUS measurements.

Fitting the parton distributions up to $x=1$ involves many complications, therefore we used the GRV98 parametrization for the large-$x$ behaviour. As pointed out previously, this does not influence the small-$x$ distributions much. Extension of this model up to $x=1$ is left for future work. Since we know that Regge models can reproduce soft-processes measurements, it should also be very interesting to check that we can also extend the form factors found in this paper down to $Q^2=0$.

Finally, we used DGLAP evolution to {\em extract} the residues from the data. As we have seen, these are smooth functions of $Q^2$ and give numerical predictions close to the DGLAP essential singularity. It should therefore be of prime interest to find an equation which stabilizes the triple-pole singularity and keeps the correct large-$Q^2$ behaviour.

\begin{acknowledgments}
I would like to thanks J.-R. Cudell for useful discussions. This work is supported by the National Fund for Scientific Research (FNRS), Belgium.
\end{acknowledgments}

\end{document}